\documentclass[11pt,eqs]{article}
\usepackage{latexsym}

\def\cleq{\setcounter{equation}{0}}

\title{
Non-geometric background arising in the solution
of Neumann boundary conditions
\thanks{Work supported in part by
the Serbian Ministry of Education and Science, under contract No. 171031.}}
\author{Lj. Davidovi\'c \thanks{e-mail: ljubica@ipb.ac.rs} and B. Sazdovi\'c
\thanks{e-mail: sazdovic@ipb.ac.rs}\\
{\it Institute of Physics,}\\
{\it University of Belgrade,}\\
{\it 11001 Belgrade, P.O.Box 57, Serbia}}

\begin{document}
\maketitle
\begin{abstract}
We investigate the open string propagation in the weakly curved background
with the Kalb-Ramond field containing
an infinitesimal part, linear in coordinate.
Solving the Neumann boundary conditions,
we find the expression for the space-time coordinates in terms of the effective ones.
So, the initial theory reduces to the effective one.
This effective theory is defined
on the non-geometric doubled space $(q^\mu,\tilde{q}_\mu)$,
where $q^\mu$ is the effective coordinate and $\tilde{q}_\mu$
is its T-dual.
The effective metric
depends on the coordinate $q^\mu$
and there exists
non-trivial effective Kalb-Ramond field which
depends on the T-dual coordinate $\tilde{q}_\mu$.
The fact that $\tilde{q}_\mu$ is $\Omega$-odd
leads to the nonvanishing effective Kalb-Ramond field.
\end{abstract}

%\PACS{
%      {11.25.-w}{Strings and branes}
%   \and
%      {11.10.Lm}{Nonlinear or nonlocal theories and models}
%     } % end of PACS codes
%}
\maketitle

%%%%%%%%%%%%%%%%%%%%%%%%%%%
%%%%%%%%%%%%%%%%%%%%%%%%%%%%%%%%%%%%%%%%%%%%%%%%%%%%%%%%%%%%%%%%%%%%%%%%%%%%%%%%%%%%%%%%%%%%%%%%%%%%%%%%%
\section{Introduction}

Recently, a formulation of the string theory in the non-geometric backgrounds
has been considered \cite{H}-\cite{BW}.
In the geometric background the transition functions between the coordinate
patches are the fundamental symmetries of the theory
as the diffeomorphisms and the gauge transformations.
But, what if the new fundamental symmetry, the T-duality is promoted to
the transition function? Then one talks about the non-geometric
T-folds and the string theory on 
such background can be consistently defined \cite{H}.

Because T-duality mixes the momentum and the winding modes,
the new transition function will relate the patches
parametrized by both the conventional coordinates $x^\mu$
(conjugate to the momenta modes)
and their T-duals $\tilde{x}^\mu$
(conjugate to the winding modes).
The theory is formulated in the doubled formalism with coordinates
$(x^\mu,\tilde{x}^\mu)$.
In order to preserve the number of degrees of freedom some relation between
$x^\mu$ and $\tilde{x}^\mu$ must be imposed.
So, while the geometric backgrounds can depend on the space-time coordinates,
the non-geometric backgrounds have the nontrivial dependence on the
dual coordinates as well \cite{H}.

Non-geometric backgrounds arise in
many different approaches.
They are often connected with the T-duality on NS-NS fluxes \cite{STW},
and compactification
with duality twists \cite{DHD}.
There also exists the formulation of the double field theory \cite{HHZ}
as the field theory on doubled torus.
More details and extended list of references can be found in the review
paper \cite{BW}.

In the present paper, we present yet another case where
the non-geometric space naturally appears.
In comparison with other investigations,
ours do not include compactified coordinates nor do
we perform the T-duality.
We find the solution of Neumann boundary conditions
on the open string endpoints,
which turns the weakly curved background of the original theory
into the doubled background of the effective theory.

We start with the theory describing the motion of the open bosonic string
in the background created by the closed string modes
$G_{\mu\nu}$ and $B_{\mu\nu}$.
In our previous papers \cite{DS} and \cite{DS1},
we treated the Neumann boundary conditions as constraints
and used the Dirac prescription.
In the present paper we show that presupposing the appropriate form of the solution,
its explicit final expression may be obtained in a much simpler way.
On this solution of boundary condition,
we obtain the effective theory.
The effective background fields 
are the background fields seen by the effective string.
In the previously investigated cases,
of the flat background ($b_{\mu\nu}\neq 0$ and $B_{\mu\nu\rho}=0$) and the weakly curved one
without constant term of Kalb-Ramond field
($b_{\mu\nu}= 0$ and $B_{\mu\nu\rho}\neq 0$ \cite{DS})
the effective metric is constant and the effective Kalb-Ramond field is zero.

In the present case of a weakly curved background
($b_{\mu\nu}\neq 0$ and $B_{\mu\nu\rho}\neq 0$)
the situation is quite different.
The effective theory is defined on the doubled target
space $(q^\mu, \tilde{q}^\mu)$.
These effective coordinates appear naturally in the solution of the Neumann boundary conditions.
Moreover, the second coordinate of the target space turns out to be the T-dual of the first one.
The effective metric depends on the effective coordinate $q^\mu$,
which is symmetric under $\sigma$-parity transformation $\Omega:\sigma\rightarrow -\sigma$ and
the effective Kalb-Ramond field depends on the ${\tilde{q}}^\nu$, which is antisymmetric.
Because of this fact
the term in the action containing the effective Kalb-Ramond field becomes $\Omega$-even,
which allows its survival.
In the conventional space, with the effective coordinate $q^\mu$,
the effective Kalb-Ramond field
comes within the $\Omega$-odd term in the action and as such is eliminated.
In the present paper,
because of the fact that the effective theory is defined in the doubled space,
the effective string can actually see the effective Kalb-Ramond field.

%%%%%%%%%%%%%%%%%%%%%%%%%%%%%%%%%%%%%%%%%%%%%%%%%%%%%%%%%%%%%%%%%%%%%%%%%%%%%%%%%%%%%%%%%%%%%%%%%%%%%%%%%%%%
\section{Open string theory in weakly curved background}
\cleq
We are considering propagation of the open bosonic string,
in the background defined by the metric tensor
$G_{\mu\nu}$ and Kalb-Ramond
antisymmetric field $B_{\mu\nu}$.
It is described by the action  \cite{GR,ZW,BBS,FTFCB,SW}
\begin{equation}\label{eq:action0}
S = \kappa \int_{\Sigma} d^2\xi\Big{[}\frac{1}{2}{\eta}^{\alpha\beta}G_{\mu\nu}(x)
+{\epsilon^{\alpha\beta}}B_{\mu\nu}(x)\Big{]}
\partial_{\alpha}x^{\mu}\partial_{\beta}x^{\nu},
\end{equation}
$(\varepsilon^{01}=-1)$, where integration goes over
two-dimensional world-sheet $\Sigma$ parametrized by the
coordinates $\xi^{0}=\tau$, $\xi^{1}=\sigma$ with $\sigma \in
[0,\pi]$. Here $x^{\mu}(\xi)$, $\mu=0,1,...,D-1$ are the
coordinates of the D-dimensional space-time, and we use the
notation $\dot{x}=\frac{\partial x}{\partial\tau}$,
$x^\prime=\frac{\partial x}{\partial\sigma}$.
The requirement for the
world-sheet conformal invariance on the quantum level,
produces the restriction on the background fields.
They must obey the space-time equations of motion.
To the lowest order in slope parameter $\alpha^\prime$, for the constant dilaton field $\Phi=const$
these equations have the form
\begin{equation}\label{eq:beta}
R_{\mu \nu} - \frac{1}{4} B_{\mu \rho \sigma}
B_{\nu}^{\ \rho \sigma}=0\, ,
\end{equation}
\begin{equation}\label{eq:beta1}
D_\rho B^{\rho}_{\ \mu \nu} = 0,
\end{equation}
where
$B_{\mu\nu\rho}=\partial_\mu B_{\nu\rho}
+\partial_\nu B_{\rho\mu}+\partial_\rho B_{\mu\nu}$
is a field strength of the field $B_{\mu \nu}$, and
$R_{\mu \nu}$ and $D_\mu$ are Ricci tensor and
covariant derivative with respect to space-time metric.
We will consider the following particular solution
of these equations,
the {\it weakly curved} background 
\cite{DS,DS1,VS,CSS}
\begin{eqnarray}\label{eq:gb}
G_{\mu\nu}=const,\quad
B_{\mu\nu}(x)=b_{\mu\nu}+\frac{1}{3}B_{\mu\nu\rho}x^\rho,
\end{eqnarray}
where the parameter $b_{\mu\nu}$ is constant and $B_{\mu\nu\rho}$
is constant and infinitesimally small.
Through the paper we will work up to the first order in $B_{\mu\nu\rho}$.

The minimal action principle for the
open string leads to the equation of motion
\begin{equation}\label{eq:motion}
{\ddot{x}}^\mu=x^{\prime\prime\mu}
-2B^\mu_{\ \nu\rho}{\dot{x}}^\nu x^{\prime\rho},
\end{equation}
and the boundary conditions on the string endpoints.
Cho\-osing the Neumann boundary conditions we have
\begin{equation}\label{eq:bonc}
\gamma^{0}_\mu\Big{|}_{\sigma=0,\pi}=0,
\end{equation}
where
\begin{equation}\label{eq:bc}
\gamma^{0}_{\mu}\equiv \frac{\delta {\cal{L}}}{\delta
x^{\prime\mu}} =G_{\mu\nu}x^{\prime\nu}-2B_{\mu\nu}\dot{x}^\nu.
\end{equation}

Our goal is
to solve the boundary conditions and obtain the
effective theory on this solution.
We achieve this, by imposing the ansatz for
the solution and by demanding
that it obeys the equations of motion and the consistency condition.
%%%%%%%%%%%%%%%%%%%%%%%%%%%%%%%%%%%%%%%%%%%%%%%%%%%%%%%%%%%%%%%%%%%%%%%%%%%%%%%%%%%%%%%%%%%%%%%%%%%%%%%%%%%%
\subsection{Solution of the boundary conditions}

It is well known that in the case of the constant background fields,
both equations of motion and boundary conditions can be solved by
expressing odd coordinate part ${\bar{q}}^\mu$
in terms of the even coordinate part ${{q}}^\mu$.
We are going to 
generalize such a solution to the case of the weakly
curved background.

Let us define the even and odd coordinate parts with respect to
$\sigma=0$
\begin{eqnarray}\label{eq:qqbar}
q^\mu(\sigma)&=&
\sum_{n=0}^{\infty}\frac{{\sigma}^{2n}}{(2n)!}x^{(2n)\mu}\Big{|}_{\sigma=0},
\nonumber\\
{\bar{q}}^\mu(\sigma)&=&
\sum_{n=0}^{\infty}\frac{{\sigma}^{2n+1}}{(2n+1)!}x^{(2n+1)\mu}\Big{|}_{\sigma=0},
\end{eqnarray}
and separate the even and odd parts of the equation of motion
(\ref{eq:motion}) and the boundary condition (\ref{eq:bonc})
at $\sigma=0$. 
In this way, the equations of motion become
\begin{eqnarray}\label{eq:motione}
{\ddot{q}}^\mu-q^{\prime\prime\mu}&=& -2B^\mu_{\ \nu\rho} \Big{[}
{\dot{q}}^\nu {\bar{q}}^{\prime\rho}+ {\dot{\bar{q}}}^\nu
q^{\prime\rho} \Big{]},
\nonumber\\
{\ddot{\bar{q}}}^\mu-{\bar{q}}^{\prime\prime\mu}&=& -2B^\mu_{\
\nu\rho} \Big{[} {\dot{q}}^\nu q^{\prime\rho} +{\dot{\bar{q}}}^\nu
{\bar{q}}^{\prime\rho} \Big{]},
\end{eqnarray}
and only the even part of $\gamma^{0}_{\mu}$ contributes to the
boundary conditions at $\sigma=0$
\begin{eqnarray} \label{eq:bcon}
&&\gamma^{0}_\mu\Big{|}_{\sigma=0}=0,
\nonumber\\
&& \gamma^{0}_{\mu}
\equiv G_{\mu\nu}{\bar{q}}^{\prime\nu} -2b_{\mu\nu}\dot{q}^\nu
-2h_{\mu\nu}(q)\dot{q}^\nu,
\end{eqnarray}
with $h_{\mu\nu}(x)=B_{\mu\nu}(x)-b_{\mu\nu}$
being infinitesimally small.

Let us look for the
solution of this boundary condition,
in the form
\begin{eqnarray}\label{eq:anzac}
\dot{\bar{q}}^\mu&=& -A^{\mu}_{1\nu} ({\tilde{q}})  \dot{q}^\nu
+2\beta^{\mu}_{1\nu} (q)q^{\prime\nu},
\nonumber\\
{\bar{q}}^{\prime\mu}&=& -A^{\mu}_{2\nu} ({\tilde{q}})
{q}^{\prime\nu} +2\beta^{\mu}_{2\nu} (q) {\dot{q}}^{\nu}.
\end{eqnarray}
We suppose that the first derivatives in $\tau$ and $\sigma$ of $\bar{q}$
are the linear functions of the first derivatives of $q$, and that they do not include the higher derivative terms.
The characteristics of the coefficients $A$ and $\beta$ arguments,
are dictated by the parity of both equations.
The coefficients $A^{\mu}_{1\nu}$ and $A^{\mu}_{2\nu}$ are odd
and as such
they do not contain constant terms and they
 depend on some odd variable ${\tilde{q}}$,
while $\beta^{\mu}_{1\nu}$ and $\beta^{\mu}_{2\nu}$ are
even functions, depending on the (new
independent) variable $q^\mu$.
Beside satisfying (\ref{eq:bcon}), the solution (\ref{eq:anzac})
must obey the equations of motion (\ref{eq:motione}),
the consistency condition $(\dot{\bar{q}}^\mu)^\prime=
({\bar{q}}^{\prime\mu})^{\cdot}$
and it must be in agreement with the zeroth order solution
\begin{eqnarray}\label{eq:zos}
\dot{\bar{q}}^\mu=2b^{\mu}_{\ \nu}q^{\prime\nu}, &&
{\bar{q}}^{\prime\mu}=2b^{\mu}_{\ \nu}{\dot{q}}^{\nu},
\end{eqnarray}
for $B_{\mu\nu\rho}=0$.

The equations of motion
(\ref{eq:motione}) and the zeroth order solution (\ref{eq:zos})
are invariant to the interchange of $\tau$ and $\sigma$
derivatives.
The boundary conditions (\ref{eq:bcon})
are also invariant because
$\gamma^{0}_{\mu}\Big{|}_{\partial_{\tau}\leftrightarrow\partial_\sigma}=
G_{\mu\nu}\dot{\bar{q}}^{\nu} -2b_{\mu\nu}{q}^{\prime\nu}
-2h_{\mu\nu}(q){q}^{\prime\nu}$ is identicaly equal to zero at $\sigma=0$.
Therefore,
we can conclude that the two ansatz equations
(\ref{eq:anzac}) must be invariant too. It follows that
$A^{\mu}_{1\nu}=A^{\mu}_{2\nu}\equiv A^{\mu}_{\ \nu}$ and
$\beta^{\mu}_{1\nu}=\beta^{\mu}_{2\nu}\equiv \beta^{\mu}_{\ \nu}$.
Moreover, comparing (\ref{eq:anzac}) with (\ref{eq:zos}), we
conclude that $A^{\mu}_{\ \nu}$ is infinitesimal.

Now, substituting (\ref{eq:anzac}) with the redefined coefficients
into the boundary condition (\ref{eq:bcon}) for $\sigma=0$, we
obtain
\begin{equation}
\Big{[}G\beta(q)\Big{]}_{\mu\nu}\Big{|}_{\sigma=0}=B_{\mu\nu}(q)\Big{|}_{\sigma=0},
\end{equation}
which gives
$\beta^{\mu}_{\ \nu}(q)=(G^{-1})^{\mu\rho}B_{\rho\nu}(q)$. The
ansatz (\ref{eq:anzac}) transforms  to
\begin{eqnarray}\label{eq:anzacc}
&&\dot{\bar{q}}^\mu= -A^{\mu}_{\ \nu}({\tilde{q}})\dot{q}^\nu
+2\Big{[}G^{-1}B(q)\Big{]}^{\mu}_{\ \nu}q^{\prime\nu}, 
\nonumber\\
&&{\bar{q}}^{\prime\mu}= -A^{\mu}_{\
\nu}({\tilde{q}}){q}^{\prime\nu}
+2\Big{[}G^{-1}B(q)\Big{]}^{\mu}_{\
\nu}{\dot{q}}^{\nu}.
\end{eqnarray}
Substituting these expressions into the equations of motion
(\ref{eq:motione}),
we obtain
\begin{equation}\label{eq:eqeven}
{\ddot{q}}^\mu-q^{\prime\prime\mu}=
12[h^{\mu}_{\
\nu}(\dot{q})b^{\nu}_{\ \rho}\dot{q}^\rho -h^{\mu}_{\
\nu}(q^\prime)b^{\nu}_{\ \rho}{q}^{\prime\rho}] ,
\end{equation}
and
\begin{eqnarray}\label{eq:ajedan}
\dot{A}^{\mu}_{\ \nu}({\tilde{q}}){\dot{q}}^\nu
-A^{\prime\mu}_{\
\nu}({\tilde{q}})q^{\prime\nu} +A^{\mu}_{\
\nu}({\tilde{q}})(\ddot{q}^\nu-q^{\prime\prime\nu}) =
2h^{\prime\mu}_{\
\nu}{\dot{q}}^\nu - 24h^{\prime\mu}_{\ \nu}(bq)(b\dot{q})^\nu,
\end{eqnarray}
and the consistency relation $(\dot{\bar{q}}^\mu)^\prime=
({\bar{q}}^{\prime\mu})^{\cdot}$ gives
\begin{equation}\label{eq:concond}
2B^{\mu}_{\ \nu}(q)({\ddot{q}}^\nu-q^{\prime\prime\nu})=
\dot{A}^{\mu}_{\ \nu}({\tilde{q}}){q}^{\prime\nu}
-A^{\prime\mu}_{\ \nu}({\tilde{q}}){\dot{q}}^{\nu}.
\end{equation}
We can eliminate
${\ddot{q}}^\nu-q^{\prime\prime\nu}$ from (\ref{eq:ajedan})
and (\ref{eq:concond}), using (\ref{eq:eqeven})
to obtain two equations
for the unknown coefficient $A^{\mu}_{\ \nu}$
\begin{eqnarray}\label{eq:ajedan1}
\dot{A}^{\mu}_{\ \nu}({\tilde{q}}){\dot{q}}^\nu -A^{\prime\mu}_{\
\nu}({\tilde{q}})q^{\prime\nu} &=&2h^{\prime\mu}_{\
\nu}{\dot{q}}^\nu - 24h^{\prime\mu}_{\ \nu}(bq)(b\dot{q})^\nu,
\end{eqnarray}
\begin{equation}\label{eq:adva}
\dot{A}^{\mu}_{\ \nu}({\tilde{q}}){q}^{\prime\nu} -A^{\prime\mu}_{\
\nu}({\tilde{q}}){\dot{q}}^{\nu} =24\Big{[}
b\dot{h}b\dot{q}-bh^{\prime}bq^\prime {\Big{]}}^\mu.
\end{equation}
Note that the third term at the left hand side of (\ref{eq:ajedan}) has
been neglected, being the higher order term as the product of the
infinitesimally small ${A}^{\mu}_{\ \nu}$ and
${\ddot{q}}^\nu-q^{\prime\prime\nu}$.

Because in all relations (equations of motions, boundary conditions and ansatz)
the sum of $\sigma$ and $\tau$ derivatives
is preserved and ${\tilde{q}}$ is odd, we conclude
\begin{eqnarray}\label{eq:dotprime}
{\dot{\tilde{q}}}^\mu= q^{\prime\mu} , \qquad
{\tilde{q}}^{\prime\mu}={\dot{q}}^\mu .
\end{eqnarray}
The interpretation of these equations will be discussed at section
\ref{sec:double}.
From (\ref{eq:dotprime}) follows
\begin{eqnarray}
\dot{A}^{\mu}_{\ \nu}({\tilde{q}})=
A^{\mu}_{\ \nu}(\dot{{\tilde{q}}})=
A^{\mu}_{\ \nu}(q^{\prime}),
&&
A^{\prime\mu}_{\ \nu}({\tilde{q}})=
A^{\mu}_{\ \nu}({\tilde{q}}^{\prime})
= A^{\mu}_{\ \nu}(\dot{q}),
\end{eqnarray}
and the solutions of (\ref{eq:ajedan1}) and (\ref{eq:adva}) are
\begin{eqnarray}
_{(1)}{A}^{\mu}_{\ \nu}({q})&=&
(G^{-1})^{\mu\rho}
\Big{[}
h({q})-12h(b{q})b
\Big{]}_{\rho\nu},
\nonumber\\
_{(2)}{A}^{\mu}_{\ \nu}({q})&=&
(G^{-1})^{\mu\rho}
\Big{[}-12
bh({q})b
+12bh(b{q})
\Big{]}_{\rho\nu}.
\end{eqnarray}

Because the solution $_{(1)}A^{\mu}_{\ \nu}$
satisfies the homogeneous part of the equation (\ref{eq:adva})
and the solution $_{(2)}A^{\mu}_{\ \nu}$
satisfies the homogeneous part of the equation (\ref{eq:ajedan1}),
the complete solution for $A^{\mu}_{\ \nu}$ is of the form
\begin{eqnarray}
A^{\mu}_{\ \nu}({q})
=(G^{-1})^{\mu\rho}
\Big{[}
h({q})-12bh({q})b
-12h(b{q})b+12bh(b{q})
\Big{]}_{\rho\nu},
\end{eqnarray}
with the property $(GA)_{\mu \nu} = - (GA)_{\nu \mu}$.
Finally, we can write the space-time coordinates satisfying
the boundary condition at $\sigma=0$ as
\begin{eqnarray}\label{eq:stcoor}
\dot{x}^\mu&=&[\delta^{\mu}_{\nu} -A^{\mu}_{\ \nu} ({\tilde{q}})]  \dot{q}^\nu
+2[G^{-1}B(q)]^{\mu}_{\ \nu} q^{\prime\nu},
\nonumber\\
x^{\prime\mu}&=&[\delta^{\mu}_{\nu} -A^{\mu}_{\ \nu} ({\tilde{q}})]
{q}^{\prime\nu} +2[G^{-1}B(q)]^{\mu}_{\ \nu}  {\dot{q}}^{\nu}.
\end{eqnarray}

For $\sigma=\pi$ instead of (\ref{eq:qqbar}), we define even and
odd variables with respect to $\sigma=\pi$
\begin{eqnarray}\label{eq:qqbarpi}
{^\star q}^\mu(\sigma)&=&
\sum_{n=0}^{\infty}\frac{{\sigma}^{2n}}{(2n)!}x^{(2n)\mu}\Big{|}_{\sigma=\pi},
\nonumber\\
{^\star{\bar{q}}}^\mu(\sigma)&=&-
\sum_{n=0}^{\infty}\frac{{\sigma}^{2n+1}}{(2n+1)!}x^{(2n+1)\mu}\Big{|}_{\sigma=\pi}.
\end{eqnarray}

Applying the analogous procedure as for the case
$\sigma=0$, we obtain
\begin{eqnarray}\label{eq:qqbarpires}
\dot{x}^\mu(\sigma)&=&
\Big{[}\delta^{\mu}_{\nu}
-A^{\mu}_{\ \nu} [{^\star{\tilde{q}}}(\pi-\sigma)]
\Big{]}
{^\star\dot{q}}^\nu(\pi-\sigma)
+2\Big{[}G^{-1}B[{^\star q}(\pi-\sigma)]\Big{]}^{\mu}_{\ \nu} {^\star
q}^{\prime\nu}(\pi-\sigma),
\nonumber\\
{x}^{\prime\mu}(\sigma)&=&
\Big{[}\delta^{\mu}_{\nu}
 -A^{\mu}_{\ \nu}
[{^\star\tilde{q}}(\pi-\sigma)]
\Big{]}
{^\star q}^{\prime\nu}(\pi-\sigma)
+2\Big{[}G^{-1}B[{^\star
q}(\pi-\sigma)]\Big{]}^{\mu}_{\ \nu} {^\star \dot{q}}^{\nu}(\pi-\sigma).
\nonumber\\
\end{eqnarray}
Note that if
\begin{equation}\label{eq:qqstar}
q^\mu(\sigma)={^\star q}^\mu(\pi-\sigma),\qquad
{\bar{q}}^\mu(\sigma)={^\star {\bar{q}}}^\mu(\pi-\sigma),
\end{equation}
then the solutions (\ref{eq:stcoor}) and (\ref{eq:qqbarpires}) coincide, and
from the relation (\ref{eq:qqstar}) follows the $2\pi$-periodicity
of $x^\mu$.
So, if we extend the $\sigma$ domain and demand $2\pi-$periodicity
of the original variable $x^\mu(\sigma+2\pi)=x^\mu(\sigma)$,
the relation (\ref{eq:stcoor}) solves both constraints at
$\sigma=0$ and $\sigma=\pi$.

Let us stress that the solution of the boundary condition does not depend on
one effective variable only, but on the two variables
$q^\mu$  and ${\tilde{q}}^\mu$, connected by the relation (\ref{eq:dotprime}).
So, we obtained some non-geometric 
space with the doubled number of degrees of freedom but with the
constraint (\ref{eq:dotprime}).
%%%%%%%%%%%%%%%%%%%%%%%%%%%%%%%%%%%%%%%%%%%%%%%%%%%%%%%%%%%%%%%%%%%%%%%%%%%%%%%%%%%%%%%%
\subsection{Effective theory}

Substituting the solution (\ref{eq:stcoor}) into the Lagrangian
(\ref{eq:action0})
we obtain the effective Lagrangian
\begin{equation}\label{eq:efflag}
{\cal{L}}^{eff}=
\frac{\kappa}{2}{\dot{q}}^{\mu}G^{eff}_{\mu\nu}(q,{\tilde{q}}){\dot{q}}^{\nu}
-\frac{\kappa}{2}{q}^{\prime\mu}G^{eff}_{\mu\nu}(q,{\tilde{q}}){q}^{\prime\nu}
+2\kappa{q}^{\prime\mu}B^{eff}_{\mu\nu}(q,{\tilde{q}}){\dot{q}}^{\nu},
\end{equation}
where
\begin{eqnarray}
G^{eff}_{\mu\nu}(q,{\tilde{q}})&=&
G^{E}_{\mu\nu}(q+2b{\tilde{q}})
+4[b^{2}A({\tilde{q}})-A({\tilde{q}})b^{2}]_{\mu\nu},
\nonumber\\
B^{eff}_{\mu\nu}(q,{\tilde{q}})&=&
[h(2b{\tilde{q}})+4bh(2b{\tilde{q}})b]_{\mu\nu}
-B_{\mu}^{\ \rho}(q)G_{\rho\nu}^{E}(q).
\end{eqnarray}
The expression
\begin{equation}
G^{E}_{\mu\nu}(x)\equiv G_{\mu\nu}-4B_{\mu\rho}(x)(G^{-1})^{\rho\sigma}B_{\sigma\nu}(x)
\end{equation}
is the definition of the open string metric,
which was introduced in \cite{SW} for the
constant background case.
In our case of the weakly curved background (in the leading order in $B_{\mu\nu\rho}$) it is equal to
\begin{eqnarray}
G^{E}_{\mu\nu}&=&g_{\mu\nu}-4(bh+hb)_{\mu\nu},
\nonumber\\ g_{\mu\nu}&=&G_{\mu\nu}-4(bG^{-1}b)_{\mu\nu}=G^{E}_{\mu\nu}(0).
\end{eqnarray}

Because our basic variable $q^\mu (\sigma)$ contains only even
powers of $\sigma$,
it is convenient to regard it as the even function
$q^\mu (-\sigma)=q^\mu (\sigma)$ on the interval $\sigma \in [-\pi,\pi]$.
Hereafter, we will consider the action
$S^{eff} = \int d\tau \int_{-\pi}^\pi d\sigma {\cal{L}}^{eff}$,
and consequently, the terms of the effective metric which depend on
${\tilde{q}}$ and the term of effective Kalb-Ramond field which depends
on $q$ will disappear,
so that
\begin{equation}\label{eq:aceff}
S^{eff} = \kappa \int_{\Sigma_1} d^2\xi   \Big[\frac{1}{2}\eta^{\alpha\beta}G^{eff}_{\mu\nu}(q)
+{\epsilon^{\alpha\beta}}B^{eff}_{\mu\nu}(2b{\tilde{q}})\Big]
\partial_{\alpha}q^{\mu}\partial_{\beta}q^{\nu}.
\end{equation}
Here $\Sigma_1$ marks the changed sigma domain $\sigma \in [-\pi, \pi]$.
The effective background fields are equal to
\begin{eqnarray}
G^{eff}_{\mu\nu}(q)&=& G^{E}_{\mu\nu}(q),
\nonumber\\
B^{eff}_{\mu\nu}(2b{\tilde{q}})&=& -\frac{\kappa}{2} [g \Delta\theta (2b{\tilde{q}})g]_{\mu\nu},
\end{eqnarray}
where $\Delta\theta$ is the infinitesimal part of
the so called non-commutativity parameter
\begin{equation}
\theta^{\mu\nu}=-\frac{2}{\kappa}\Big{[}G^{-1}_{E}BG^{-1}\Big{]}^{\mu\nu}
=\theta_{0}^{\mu\nu}
-\frac{2}{\kappa}\Big{[}g^{-1}(h+4bhb)g^{-1}\Big{]}^{\mu\nu}.
\end{equation}
It is defined in analogy with that of the flat space-time
introduced in \cite{SW}.
The constant parts of the
effective metric
and the non-commutativity parameter
are denoted by
$g_{\mu\nu}=G^{E}_{\mu\nu}(0)$
and $\theta_{0}^{\mu\nu}=\theta^{\mu\nu}(0)=
-\frac{2}{\kappa}\Big{[}g^{-1}bG^{-1}\Big{]}^{\mu\nu}$.

%%%%%%%%%%%%%%%%%%%%%%%%%%%%%%%%%%%%%%%%%%%%%%%%%%%%%%%%%%%%%%%%%%%%%%%
\section{Doubled geometry of the effective theory}\label{sec:double}
\cleq
There are two somewhat unexpected
things in the effective theory (\ref{eq:aceff}).
The first one is the appearance
of the non-trivial Kalb-Ramond field $B^{eff}_{\mu\nu}$.
The second one, which is essentially the cause of the first,
is the fact that it does not depend on the coordinate
$q^\mu$ but on ${\tilde{q}}^\mu$.
Let us analyze these results
pursuing an explanation and interpretation.

It is well known that the theory of the unoriented closed string,
which is for example the effective theory for the constant background, 
does not contain the Kalb-Ramond field.
Standardly, this is explained \cite{ZW,SN} by noticing
that the effective Kalb-Ramond field appears in the
effective action within the term
$B^{eff}_{\mu\nu}{\dot{q}}^\mu q^{\prime\nu}$.
If the Kalb-Ramond field depends on the $\Omega$-even variable,
this term does not contribute.

In the case of the weakly curved background
the effective Kalb-Ramond field does not depend
on the effective coordinate $q^\mu$ ($\Omega$-even)
but on
${\tilde{q}}^\mu$ ($\Omega$-odd).
Since, $B_{\mu\nu}^{eff}(2b{\tilde{q}})$
is proportional to ${\tilde{q}}^\mu$
it means that the effective Kalb-Ramond field
is odd under $\sigma$-parity transformation
$
\Omega B_{\mu\nu}^{eff}[2b{\tilde{q}}(\sigma)]=-B_{\mu\nu}^{eff}[2b{\tilde{q}}(\sigma)].
$
This makes the term
$B^{eff}_{\mu\nu}(2b{\tilde{q}}){\dot{q}}^\mu q^{\prime\nu}$
$\Omega$-even and allows its survival.

To find the interpretation of ${\tilde{q}}^\mu$,
let us solve the relations in (\ref{eq:dotprime}).
First, note that ${\tilde{q}}^\mu$
appears as an argument of $B^{eff}_{\mu\nu}$ only, which is the
infinitesimal of the first order. So, it is enough to consider
${\tilde{q}}^\mu$ in the zeroth order. The zeroth order equation of
motion for $q^\mu$ is just $\partial_{+}\partial_{-}q^\mu=0$, and
consequently the solution has the form
$q^\mu(\sigma)=f^\mu(\sigma^{+})+g^\mu(\sigma^{-})$,
with $\sigma^{\pm}=\tau\pm\sigma$.
The $\Omega$-evenness of the variable
$q^\mu$,
$q^\mu(-\sigma)=q^\mu(\sigma)$,
implies $f(\sigma)=g(\sigma)$
and we obtain
\begin{equation}
q^\mu(\sigma)=f^\mu(\sigma^{+})+f^\mu(\sigma^{-}).
\end{equation}

From the properties $\partial_{\pm}f^{\mu}(\sigma^{\mp})=0$, we have
${\dot{f}}^{\mu}(\sigma^{\pm})=\pm f^{\prime\mu}(\sigma^{\pm})$.
Therefore, $ {\dot{q}}^\mu(\sigma)=f^{\prime\mu}(\sigma^{+})
-f^{\prime\mu}(\sigma^{-}), $ and consequently for both equations
from (\ref{eq:dotprime}) we obtain
\begin{equation}\label{eq:qtil}
{\tilde{q}}^\mu(\tau,\sigma)= f^{\mu}(\sigma^{+})
-f^{\mu}(\sigma^{-}) +f_{0}.
\end{equation}
The variable ${\tilde{q}}^\mu$ is odd by definition,
so it can not contain the constant part, therefore
the integration constant $f_{0}$ in (\ref{eq:qtil}) is zero.
This means that
${\tilde{q}}^\mu(\tau,\sigma)$ is T-dual mapping of the
effective coordinate $q^\mu(\tau,\sigma)$ (see for example
(17.76) of Ref. \cite{ZW}
or eq. (6.17) of Ref.\cite{BBS}).

%%%%%%%%%%%%%%%%%%%%%%%%%%%%%%%%%%%
Let us comment on the above conclusion.
The Buscher construction of the T-dual theory,
done for the constant background,
can be generalized to the case
of the weakly curved background Ref. \cite{DS2}.
The dual coordinate $y_\mu$ can be expressed in terms of the
original one $x^\mu$ as
\begin{equation}\label{eq:ctl}
\partial_{\pm}y_\mu\cong
-2\Pi_{\mp\mu\nu}[x]\partial_{\pm}x^\nu\mp
2\beta^{\mp}_{\mu}[x],
\end{equation}
where $\Pi_{\pm\mu\nu}\equiv B_{\mu\nu}\pm\frac{1}{2}G_{\mu\nu}
$ and $\beta^{\alpha}_{\mu}[x]\equiv
\partial_\mu B_{\nu\rho}\epsilon^{\alpha\beta}x^{\nu}
\partial_{\beta}x^{\rho}$ is infinitesimal.
In the present article,
we discuss the T-dual coordinate of
$q^\mu$, the even part of coordinate $x^\mu$.
So, the role of the
initial coordinate $x^\mu$ takes $q^\mu$,
and the role of the dual coordinate $y_\mu$ takes $\tilde{q}^\nu$.
Projecting (\ref{eq:ctl}) into odd and even part and
neglecting first order in $B_{\mu\nu\rho}$ terms,
because the dual coordinate 
is the argument of the infinitesimally
small $A^\mu_{\ \nu}$,
one obtains
\begin{equation}
\dot{\tilde{q}}^\mu\cong q^{\prime\mu},
\quad
\tilde{q}^{\prime\mu}\cong \dot{q}^\mu.
\end{equation}
These are just the expressions (\ref{eq:dotprime}).
So, $q^\mu$ and $\tilde{q}^\nu$
are not just duals in the zeroth order 
but truly the dual coordinates.

%%%%%%%%%%%%%%%%%%%%%%%%%%%%%%%

So, in the effective theory,
the effective metric depends on the effective coordinate
$q^\mu$ and the effective Kalb-Ra\-mond field on its T-dual
$\tilde{q}^\mu$.
This kind of background is seen to be possible in
the so called doubled formulation
(the analysis of the nongeometrical backgrounds where the T-duality
is allowed as the transition function).
The relations (\ref{eq:dotprime}) we just solved,
represent the self duality constraint in terminology of Ref. \cite{H}
which reduces the number of the degrees of freedom by half
in the doubled non-geometrical space of $q^\mu$ and $\tilde{q}^\mu$.

%%%%%%%%%%%%%%%%%%%%%%%%%%%%%%%%%%%%%
\section{The effective Kalb-Ramond field as a torsi\-on potential}
\cleq

The effective theory (\ref{eq:aceff}) is defined on the nongeometrical space
$(q^\mu, \tilde{q}^\mu)$ where the second coordinate ${\tilde{q}}^\mu$  is
the T-dual of the first effective coordinate
$q^\mu$.
It is interesting to find the equation of motion
of the theory of this type,
when $G^{eff}_{\mu\nu}$ is arbitrary function of $q^\mu$ and
$B^{eff}_{\mu\nu}$ consists of the odd powers in $\Omega$-odd
variable $\bar{q}=2b{\tilde{q}}$.

Even-though the effective theory is defined in the doubled space with coordinates
$q^\mu$ and $\tilde{q}^\mu$,
these are related by the self-duality constraint (\ref{eq:dotprime}).
So, to obtain the equations of motion, we will vary the action with respect to the
independent variable $q^\mu$.
They contain the non-local part which is obtained by varying with respect to $\tilde{q}$.
The form of non-local term is a consequence of the relation
$\delta\dot{q}=\delta\tilde{q}^\prime$,
which follows from (\ref{eq:dotprime}).

The variation produces
\begin{eqnarray}
\eta^{\alpha\beta}\ {}^{0}D_{\beta}\partial_\alpha q^\mu
=2\kappa \theta^{\mu\nu}_{0}
\epsilon^{\alpha\beta}\int_{0}^{\sigma}d\eta
\frac{\partial}{\partial\tau}
\Big{[}
\frac{\partial B^{eff}_{\rho\sigma}(\bar{q})}{\partial\bar{q}^{\nu}}
\partial_{\alpha}q^\rho\partial_{\beta}q^\sigma
\Big{]},
\end{eqnarray}
where
${}^{0}D_{\alpha}V^\mu=\partial_\alpha q^\nu {}^{0}D_{\nu}V^\mu$
is the covariant derivative in the world-sheet direction.
${ }^{0}D_\mu$ is the generalized covariant derivative
$^{0}D_\mu V^\nu=\partial_\mu V^\nu+
{ }^{0}\Gamma^{\nu}_{\rho\mu}V^\rho$,
with the generalized connection
\begin{equation}
{ }^{0}\Gamma^{\mu}_{\rho\sigma}=(\Gamma_{eff})^{\mu}_{\rho\sigma}
+\frac{1}{2}{ }^{0}K^\mu_{\ \rho\sigma}.
\end{equation}
It consists of the Christoffel connection $(\Gamma_{eff})^{\mu}_{\rho\sigma}$
and contorsion ${ }^{0}K_{\mu\rho\sigma}=\frac{1}{2}{ }^{0}T_{\{\sigma\mu\rho\}}$
in terms of torsion
\begin{equation}\label{eq:qstr}
{ }^{0}T_{\mu\rho\sigma}=
4b_{\rho}^{\ \nu}
\frac{\partial}{\partial {\bar{q}}^\nu}
 B^{eff}_{\mu\sigma}.
\end{equation}
Here, $\{\mu\nu\rho\}=\nu\rho\mu+\rho\mu\nu-\mu\nu\rho$
is Schouten bracket.

Therefore, the Kalb-Ramond field is related to the
torsion potential.
This
is in accordance
with the
usual interpretation
of the Kalb-Ramond field $B_{\mu\nu}$
in the low energy string theory
as the non-Rimannian theory \cite{SP}.
In our case torsion is an
infinitesimally small constant.
%%%%%%%%%%%%%%%%%%%%%%%%%%%%%%%%%%%%%%%%%%%%%%%%%%%%%%%%%%%%%%%%%%%%%%%%%%%%%%%%%%%%%%%%%%%%%%
\section{Conclusion}
\cleq

In the present paper we demonstrated
that the solution of the Neumann boundary conditions
on the open string endpoints
in the weakly curved background,
naturally leads to the effective theory in the non-geometric background.
This background depends
both on the conventional effective coordinate $q^\mu$
and on its T-dual $\tilde{q}^\mu$.

We started with the oriented open string theory,
and solved the Neumann boundary conditions.
We presupposed the form of the solution
and found its explicit expression.
On the solution of the boundary conditions,
the theory reduced to the effective theory.
In the
case of the constant background fields,
the effective theory turns to be a closed string theory
defined on the unoriented orbifold,
because the effective coordinate is symmetric under $\sigma$-parity
transformation, $\Omega :\sigma\rightarrow -\sigma$ and satisfies the boundary condition
$q^\mu(\sigma=-\pi)=q^\mu(\sigma=\pi)$.

In our case, the complete transition from the original theory
(\ref{eq:action0}) to the effective theory (\ref{eq:aceff})
consists of
\begin{itemize}
\item[1.]{ the transition from conventional to the {\it doubled geometry}
\begin{equation}\label{eq:vt}
x^\mu\rightarrow q^\mu,\, {\tilde{q}}^\mu
\end{equation}
}
 \item[2.]{ {\it the background field} transition
\begin{eqnarray}\label{eq:ggbb}
G_{\mu\nu}\rightarrow G_{\mu\nu}^{eff}(q), \qquad
B_{\mu\nu}(x)\rightarrow B^{eff}_{\mu\nu}(2b{\tilde{q}}) \, .
\end{eqnarray}
}
\end{itemize}

The first transition says that
while the original theory is defined on the geometric target space,
the effective one is defined on the enlarged, the so called doubled target space,
given in terms of both the effective coordinate $q^\mu$ and its T-dual $\tilde{q}^\mu$.
Such a space is  non-geometrical space \cite{H}. 
In our case of the weakly curved background,
the non-geometrical space arose naturally in the solution
of the Neumann boundary conditions.

The appearance of the doubled target space allowed the string to see
the effective background field $B^{eff}_{\mu\nu}$.
In fact, the effective theory is $\Omega$-even projection of the initial one,
therefore in the geometric background the term with Kalb-Ramond field vanished as the $\Omega$-odd
term in the action.
But, in doubled target space Kalb-Ramond field depends on the T-dual coordinate $\tilde{q}^\mu$,
which is $\Omega$-odd.
So, the corresponding term in the action becomes $\Omega$-even and can not be projected out by the world-sheet
parity projection.

Let us summarize the forms of the initial 
and the corresponding effective
backgrounds in the following table.\\

\begin{tabular}{ll}
\hline\hline\noalign{\smallskip}
{\it Original metric} $G_{\mu\nu}$  &{\it Original KR field} $B_{\mu\nu}$\\
{\it Effective metric} $G^{eff}_{\mu\nu}$ &{\it Effective KR field} $B^{eff}_{\mu\nu}$\\
\hline\hline\noalign{\smallskip}
$G_{\mu\nu}$ & $b_{\mu\nu}$ 
\\
$g_{\mu\nu}=G_{\mu\nu}-4b_{\mu\rho}b^{\rho}_{\ \nu}$ & $0$\\
\noalign{\smallskip}\hline\noalign{\smallskip}
$G_{\mu\nu}$ & $\frac{1}{3}B_{\mu\nu\rho}x^\rho$
\\
$G_{\mu\nu}$ & $0$\\
\noalign{\smallskip}\hline\noalign{\smallskip}
$G_{\mu\nu}$ & $b_{\mu\nu}+\frac{1}{3}B_{\mu\nu\rho}x^\rho$
\\
$G_{\mu\nu}-4B_{\mu\rho}(q)
B^{\rho}_{\ \nu}(q)$ & $-\frac{\kappa}{2}(g\Delta\theta(2b\tilde{q}) g)_{\mu\nu}$\\
\noalign{\smallskip}\hline\hline\noalign{\smallskip}
\end{tabular}\\

The doubled target space and nontrivial Kalb-Ramond field appear
only in the
third row for the weakly curved background of the original theory.

In contrast to the great majority of papers
where the flat constant background is assumed,
in the present paper we investigated the string in the
coordinate dependent background.
Unexpectedly,
the effective theory obtained in this case,
significantly differs from the one obtained in the flat background.
Here,
not only that the effective background is not constant,
it can not be described by the single effective coordinate $q^\mu$,
but one needs the additional variable $\tilde{q}^\mu$.
So, considering the problem of solving the boundary conditions
in the curved background,
the doubled space appears naturally.
We also derived the equation of motion in the doubled space,
which is nontrivial because variables $q^\mu$ and $\tilde{q}^\mu$
are not independent.


\begin{thebibliography}{99}

\bibitem{H} C. Hull, JHEP \textbf{10} (2005) 065;
A. Dabholkar and C. Hull, JHEP \textbf{05} (2006) 009;
C. Hull, JHEP \textbf{07} (2007) 080.

\bibitem{STW} J. Shelton, W. Taylor and B. Wecht, JHEP \textbf{10} (2005) 085.

\bibitem{DHD} A. Dabholkar and C. Hull, JHEP \textbf{09} (2003) 054;
A. Flournoy and B. Williams, JHEP \textbf{01} (2006) 166.

\bibitem{HHZ} T. Kugo, B. Zwiebach, Prog. Theor. Phys. \textbf{87} (1992) 801;
O. Hohm, C. Hull and B. Zwiebach, JHEP \textbf{08} (2010) 008;
O. Hohm, C. Hull and B. Zwiebach, JHEP \textbf{07} (2010) 016.

\bibitem{BW} B. Wecht, Class.Quant.Grav. \textbf{24} {\bf 21} (2007) S773.

\bibitem{DS} Lj. Davidovi\'c and B. Sazdovi\'c, Phys. Rev. \textbf{D 83} (2011) 066014.

\bibitem{DS1} Lj. Davidovi\'c and B. Sazdovi\'c, JHEP \textbf{08} (2011) 112.


\bibitem{GR} M. B. Green, J. H. Schwarz and E. Witten,
\textit{Superstring
Theory}, (Cambridge University Press, 1987);
J. Polchinski, \textit{String theory}, (Cambridge University Press, 1998).

\bibitem{ZW} Zwiebach, \textit{A First Course in String Theory}, (Cambridge University Press, 2004).

\bibitem{BBS} K. Backer, M. Backer and J. Schwarz, \textit{String Theory and M-theory}, (Cambridge University Press, 2007);
C.V. Johnson, \textit{D-branes}, (Cambridge University Press, 2003).

\bibitem{FTFCB} E. S. Fradkin and A. A. Tseytlin, Phys. Lett. \textbf{B 158} (1985) 316;
Nucl. Phys. \textbf{B 261} (1985) 1;
C. G. Callan, D. Friedan, E. J. Martinec and M. J. Perry, Nucl. Phys.
\textbf{B 262} (1985) 593;
T. Banks, D. Nemeschansky and A. Sen, Nucl. Phys. \textbf{B 277} (1986) 67.

\bibitem{SW} N. Seiberg and E. Witten, JHEP \textbf{09} (1999) 032.

\bibitem{VS} V.Schomerus, {\sl Lectures of Branes in Curved Backgrounds},
Class. Quant. Grav. \textbf{19} (2002) 5781.

\bibitem{CSS} L.Cornalba and R.Schiappa, Commun.Math.Phys. \textbf{225} (2002) 33;
M. Herbst, A. Kling and M. Kreuzer, JHEP \textbf{09} (2001) 014;
A.Yu. Alekseev, A. Recknagel and V. Schomerus, JHEP \textbf{09} (1999) 023.

\bibitem{SN} B. Sazdovi\'c, Eur. Phys. J \textbf{C44} (2005) 599; 
B. Nikoli\'c and B. Sazdovi\'c, Phys. Rev. \textbf{D74} (2006) 045024;
B. Nikoli\'c and B. Sazdovi\'c, Phys. Rev. \textbf{D75} (2007) 085011;
B. Nikoli\'c and B. Sazdovi\'c, Adv. Theor. Math. Phys. \textbf{14} (2010) 1.

\bibitem{SP} T.L.Curtright and C.K.Zachos, Phys. Rev. Lett. \textbf{ 53} (1984) 1799;
Phys. Rev. \textbf{B 162}, (1985) 345;
D.S.Popovi\'c and B. Sazdovi\'c, Eur. Phys. J \textbf{C 50} (2007) 683.

\bibitem{DS2} Lj. Davidovi\'c and B. Sazdovi\'c,
\textit{T-duality in weakly curved background}, hep-th/{1205.1991}.
\end{thebibliography}
\end{document}